\documentclass[aps,prl,reprint,superscriptaddress,10pt]{revtex4-2}
\usepackage[T1]{fontenc}

\usepackage{graphicx,amsmath,amssymb,dcolumn,physics,color}

\begin{document}

\title{Decomposing Non-Markovian History Dependence}

\author{Matthew P.~Leighton}
\email{matthew.leighton@yale.edu}
\affiliation{Department of Physics and Quantitative Biology
Institute, Yale University, New Haven, CT 06511}

\author{Christopher W.~Lynn}
\email{christopher.lynn@yale.edu}
\affiliation{Department of Physics and Quantitative Biology
Institute, Yale University, New Haven, CT 06511}
\affiliation{Wu Tsai Institute, Yale University, New Haven, CT 06510}

\begin{abstract}
Non-Markovian stochastic processes are ubiquitous in biology. Nevertheless, we lack a general framework for quantifying historical dependencies. In this Letter, we propose an information-theoretic approach to decompose history dependence in systems with non-Markovian dynamics, quantifying the information encoded in dependencies of each order. In minimal models of non-Markovian dynamics, we show that this framework correctly captures the underlying historical dependencies, even when autocorrelations do not. In prolonged recordings of fly behavior, we find that the scaling of non-Markovian dependencies is invariant across timescales from fractions of a second to minutes. Despite this invariance, the overall amount of non-Markovian information is non-monotonic, suggesting a unique timescale on which historical dependencies are strongest.
\end{abstract}

\maketitle

Physics, as described by its fundamental laws, is Markovian: the next state of a closed classical or quantum system depends only on its current state. Biology, however, is generally not: dynamics can depend on states far into the past. These non-Markovian dependencies emerge as a consequence of coarse-grained descriptions, which are necessary to make sense of living systems without complete knowledge of their physical details~\cite{agon2018coarse,strasberg2019non,schwarz2024mind}. As a result, non-Markovian dynamics characterize biological systems across scales, from molecular dynamics~\cite{vroylandt2022likelihood}, epigenetic memory~\cite{d2014mechanisms}, and neural activity~\cite{zeraati2023intrinsic, cavanagh2020diversity} to organismal growth~\cite{elgamel2023multigenerational}, behavior~\cite{alba2020exploring, bialek2024long, barabasi2005origin}, and communication~\cite{shannon1948mathematical, wieczynski2025long}. 

Despite the ubiquity of non-Markovian dynamics, we lack a principled framework for quantifying the strengths of historical dependencies. Non-Markovian processes can be categorized based on their order---their longest dependence---but this provides no information about the strengths of these dependencies~\cite{iosifescu2014finite}. 
Meanwhile, autocorrelations provide key insights into a wide range of biological phenomena, including neural activity~\cite{zeraati2023intrinsic, cavanagh2020diversity, linkenkaer2001long}, animal behavior~\cite{berman2016predictability, alba2020exploring, bialek2024long, stephens2008dimensionality, wiltschko2015mapping, barabasi2005origin}, and ecology~\cite{koenig1999spatial}. However, autocorrelations can fail to characterize the underlying dependencies, even qualitatively. More advanced measures have been developed to improve upon correlations~\cite{strelioff2007inferring, james2014many, riechers2018spectral, jurgens2021shannon, radaelli2023fisher, riechers2023ultimate}, but basic questions remain: How strongly do a system's dynamics depend on the past, and can it be decomposed into simpler parts?

Answering these questions is fundamentally about prediction: Knowledge of the past reduces our uncertainty about the future, with stronger dependencies leading to larger reductions. In this Letter, we use information theory to quantify this reduction in uncertainty, which we refer to as \textit{dynamical information}. We demonstrate that dynamical information, which quantifies the total strength of historical dependencies, naturally decomposes into a term arising from the Markovian dependence on the previous state, and a series of non-negative terms contributed by each order of the non-Markovian dynamics. In minimal non-Markovian models, this decomposition captures, quantitatively, the true underlying dependencies. In large-scale recordings of fruit fly behavior---which exhibits long-range correlations~\cite{bialek2024long, branson2009high, berman2014mapping, berman2016predictability}---we discover historical dependencies that are invariant across multiple timescales. Together, these results present a principled framework for understanding the origins of non-Markovian dynamics in living systems.

\begin{figure}[t]
    \centering
    \includegraphics[width =\linewidth]{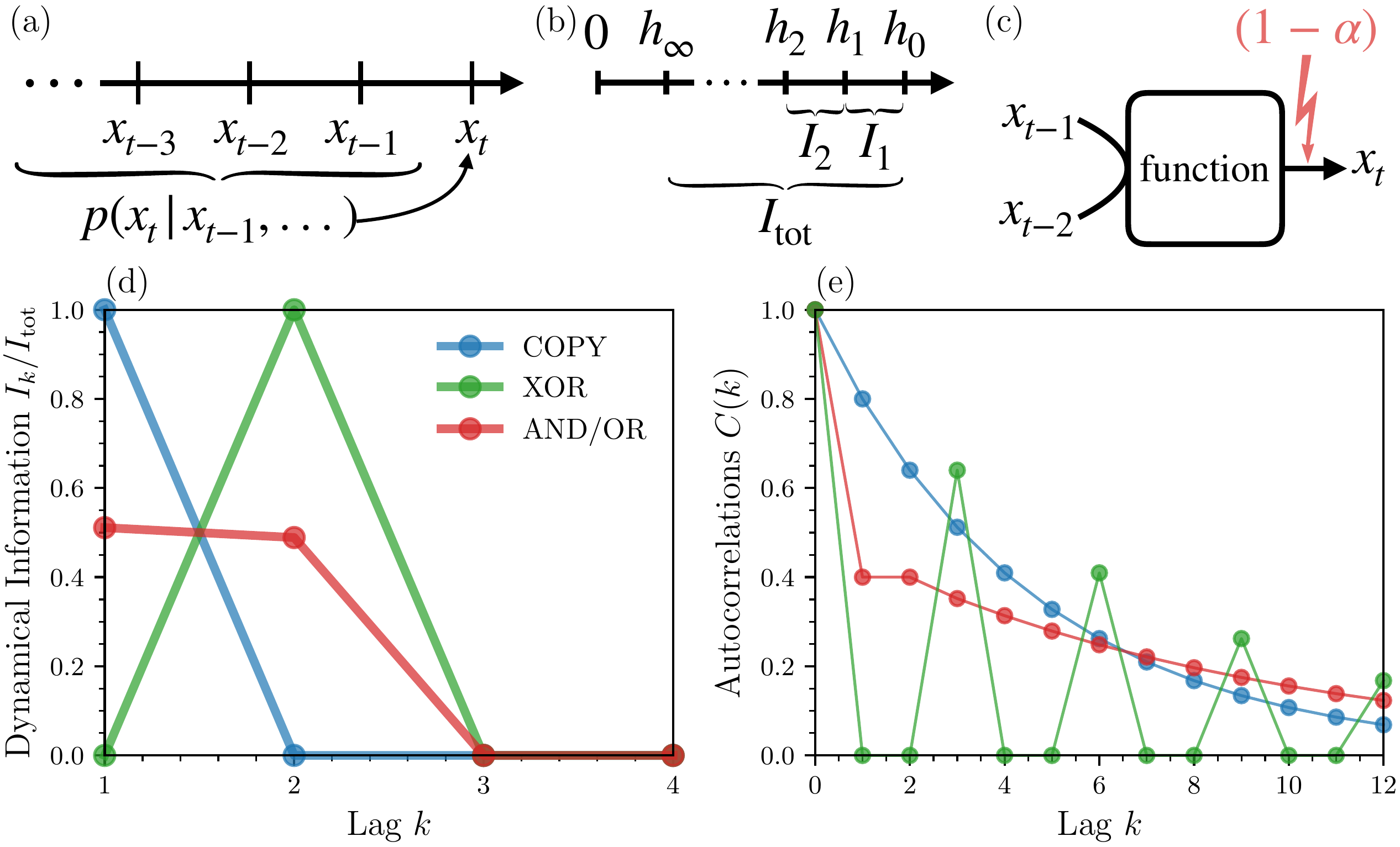}
    \caption{Decomposing history dependence in simple stochastic processes. (a) Non-Markovian dynamics are defined by the conditional distribution $p(x_t|x_{t-1},...)$. (b) Increasing knowledge of the past leads to a hierarchy of entropies $h_k$. The total dynamical information $I_\text{tot}$ therefore decomposes into a sum of non-negative contributions $I_k$ from different orders $k$. (c) Schematic of minimal dynamics in which the binary state $x_t$ performs a logical function on $x_{t-1}$ and $x_{t-2}$ with probability of success $\alpha$. (d-e) Dynamical information $I_k$ normalized by the total information $I_\text{tot}$ (d) and autocorrelations $C(k)$ (e) for different logical functions at steady-state. We set $\alpha=0.9$.}
    \label{fig:fig0}
\end{figure}

\emph{Dynamical information}.---Consider a system whose dynamics evolve in discrete steps with state $x_t$ at time $t$. The dynamics are defined by the conditional probability $p(x_t|x_{t-1},...)$, which, in general, can depend on the entire history of the system [Fig.~\ref{fig:fig0}(a)]. If the current state depends only on the previous state, such that $p(x_t|x_{t-1},...) = p(x_t|x_{t-1})$, then the dynamics are Markovian. If instead each state depends only on the previous $N$ states, then $p(x_t|x_{t-1},...) = p(x_t|x_{t-1},...,x_{t-N})$, and the dynamics are of order $N$.

To quantify the strength of these dependencies, consider the problem of predicting the next state $x_t$. With no knowledge of the past, our uncertainty is defined by the marginal entropy
\begin{equation}
h_0 \equiv H[x_t] = - \langle \log p(x_t) \rangle,
\end{equation}
where angle brackets denote an average over $p(x_t,x_{t-1},...)$. By contrast, if we have access to the entire history of the system, then our uncertainty is defined by the entropy rate
\begin{equation}
h_\infty \equiv H[x_t|x_{t-1},\hdots] = - \langle \log p(x_t | x_{t-1},\hdots) \rangle,
\end{equation}
which quantifies the inherent stochasticity of the dynamics, even with full knowledge of the past~\cite{shannon1948mathematical}. Since entropy can only decrease under conditioning~\cite{Cover2006_Elements}, knowing the past can only reduce our uncertainty, such that $h_0 \ge h_\infty$. This reduction in entropy is precisely the information that the past carries about the next state,
\begin{equation}
I_\text{tot} \equiv h_0 - h_\infty,
\end{equation}
which we refer to as the \textit{dynamical information}.

The dynamical information $I_\text{tot}$ (distinct from predictive information~\cite{bialek2001predictability}) is equivalently the mutual information between the next state of a system and its entire history. Moreover, $I_\mathrm{tot}$ is also the rate of growth of the total correlation along the trajectory (see Appendix).

Between the extremes of no knowledge and full knowledge of the past, one could have partial access to the history of the system. Given the previous $k$ states, our uncertainty about $x_t$ is given by the conditional entropy,
\begin{equation}
\begin{aligned}
h_k &\equiv H[x_t|x_{t-1},\hdots,x_{t-k}], \\
\end{aligned}
\end{equation}
which has previously been called the myopic information rate~\cite{riechers2018spectral} or differential entropy~\cite{melnik2014entropy}. As $k$ increases and we gain more knowledge of the past, this uncertainty can only decrease, resulting in a hierarchy of entropies,
\begin{equation}
\label{eq_h}
h_0 \ge h_1 \ge \cdots \ge h_k \ge \cdots \ge h_\infty \ge 0.
\end{equation}

To understand the origins of dynamical information, it is natural to compare $h_k$ and $h_{k-1}$. If $h_k = h_{k-1}$, then the state $k$ steps in the past is redundant in the sense that the dynamics are entirely determined by lower-order dependencies. By contrast, if $h_k < h_{k-1}$, then the $k^\text{th}$-order dependence provides new information about the future of the system. In this way, the strength of the $k^\text{th}$-order dependence is quantified by the difference
\begin{equation}
I_k \equiv h_{k-1} - h_k \ge 0,
\end{equation}
which we refer to as the \textit{$k^\text{th}$-order dynamical information}. This is equivalent to the mutual information between $x_t$ and $x_{t-k}$ conditioned on the intervening history,
\begin{equation}
I_k = I[x_t;x_{t-k}\, |\, x_{t-1},\hdots,x_{t-k+1}].
\end{equation}
Summing over all orders, we arrive at a full decomposition of the dynamical information [Fig.~\ref{fig:fig0}(b)]:
\begin{equation}
I_\text{tot} = I_1 + I_2 + \cdots = \sum_{k = 1}^\infty I_k.
\end{equation}
This is our main theoretical result, decomposing the total strength of non-Markovian dependencies into non-negative contributions from each order $k$.

\emph{Finite-order dynamics}.---As an important case, consider dynamics of finite order $N$. States beyond $N$ steps in the past do not reduce our uncertainty about $x_t$, so that $h_N = h_{N+1} = \cdots = h_\infty$. The dynamical information vanishes ($I_k = 0$) for all orders $k > N$, and thus recovers the correct order of the underlying dynamics.

As a simple example, consider a minimal process in which, at each point in time, the binary state $x_t$ is a noisy logical function of the previous two states $x_{t-1}$ and $x_{t-2}$ [Fig.~\ref{fig:fig0}(c)]. Since these dynamics are second-order, only $I_1$ and $I_2$ contribute to the dynamical information. The simplest function copies the previous state $x_{t-1}$ while ignoring $x_{t-2}$. In this case, the second-order information vanishes ($I_2 = 0$), and thus the history dependence is entirely Markovian ($I_\text{tot} = I_1$) as expected [Fig.~\ref{fig:fig0}(d)]. For the functions AND and OR, the dynamics are more complex, with dynamical information divided almost evenly between first and second order. Indeed, for both AND and OR, $x_t$ tends to increase with $x_{t-1}$ (yielding $I_1 > 0$), yet the full dynamics are not defined until both orders are taken into account (yielding $I_2 > 0$). Meanwhile, XOR defines a purely irreducible dependence, with the previous state alone providing no information. As such, the Markovian information vanishes ($I_1 = 0$), and the dynamics are entirely second order [$I_\text{tot} = I_2$; Fig.~\ref{fig:fig0}(d)].

\begin{figure*}[t]
    \centering
    \includegraphics[width =\linewidth]{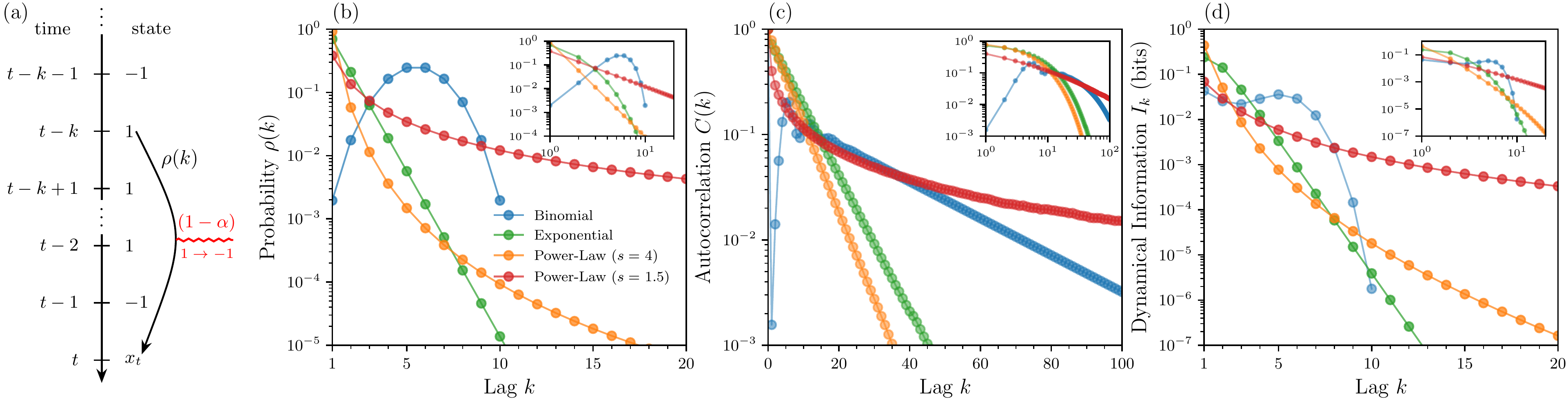}
    \caption{Autocorrelations and dynamical information in a non-Markovian copying process. (a) At each step $t$, the state $k$ steps back is selected with probability $\rho(k)$ and then copied with probability $\alpha$~\cite{leighton2025companion}. (b) Different history dependence distributions $\rho(k)$, including binomial [$\rho(k)={N-1\choose k-1}q^{k-1}(1-q)^{N-k}$ with $N=10,q=0.5$], exponential [$\rho(k)\propto (1-q)^{k-1}$ with $q=0.7$], and power-law [$\rho(k)\propto k^{-s}$ with $s=4,1.5$]. (c-d) Autocorrelations $C(k)$ (c) and dynamical information $I_k$ (d) for copying processes with different history dependencies $\rho(k)$. In all panels, insets display log-log scales and dynamics are defined with $\alpha=0.9$.}
    \label{fig:fig2}
\end{figure*}

By contrast, finite-order processes produce autocorrelations $C(k) = \langle x_t x_{t-k} \rangle - \langle x_t \rangle\langle x_{t-k} \rangle$ at all orders, making it difficult to discern the true order of dynamics [Fig.~\ref{fig:fig0}(e)]. Moreover, autocorrelations only detect pairwise dependencies between states, and therefore are fundamentally insufficient to capture complex higher-order dependencies. For example, the XOR function produces no correlations up to second order (where the true dependencies lie) but generates spurious correlations at multiples of $k = 3$ [Fig.~\ref{fig:fig0}(e)]. Thus, while autocorrelations can obscure non-Markovian dependencies (even in relatively simple processes), the strengths of these dependencies can still be quantified using the dynamical information.

\emph{Quantifying long-range dependencies}.---When confronted with biological data, we generally expect non-vanishing dependencies of all orders. The nature of these dependencies can often be characterized by their scaling with $k$: either exponential (reflecting short-range dependencies) or power-law (long-range). It is tempting to expect the large-$k$ scaling of autocorrelations $C(k)$ to match that of the dependencies. While correct in many cases, there are nonetheless many counterexamples that defy this expectation. 
Can the dynamical information recover the correct scaling even when autocorrelations do not?

Among the most striking examples, some Markovian processes, which dynamical information correctly identifies as first order ($I_\text{tot} = I_1$), can produce power-law autocorrelations. These include diffusion in logarithmic potential energy landscapes~\cite{micciche2009modeling,lillo2002long,dechant2012superaging} or with multiplicative noise~\cite{ruseckas20101}, and other exotic stochastic processes~\cite{srokowski2004stochastic, ruseckas20101}. As another example, Markovian processes with many degrees of freedom can exhibit power-law autocorrelations near criticality~\cite{godreche2000response,ramasco2004ageing,chatelain2004universality,pleimling2005dynamic,albano2011study,venturelli2022nonequilibrium}. This scenario is particularly relevant for biological systems, where increasing evidence hints at the prevalence of critical phenomena~\cite{mora2011biological,munoz2018colloquium}. 

Conversely, and perhaps more surprisingly, non-Markovian processes with truly long-range dependencies can nonetheless exhibit exponentially-decaying autocorrelations. To gain intuition for the interplay between history dependence, dynamical information, and autocorrelations, we consider a class of analytically tractable non-Markovian dynamics with dependencies that can be tuned directly. We highlight key results here and point the reader to our companion paper for full details~\cite{leighton2025companion}.

Consider a discrete-time system with states $x_t = \pm 1$. At each step $t$, the system looks into the past and selects the state $k$ steps back with probability $\rho(k)$. The next state $x_t$ then successfully copies $x_{t-k}$ with probability $\alpha\in[0,1]$ or fails (resulting in $x_t = -x_{t-k}$) with probability $1-\alpha$ [Fig.~\ref{fig:fig2}(a)]. The distribution $\rho(k)$ encodes the strength of non-Markovian dependencies [Fig.~\ref{fig:fig2}(b)]. For example, if $\rho(k)$ is a delta function centered at $k = 1$, then we recover the copy function from Fig.~\ref{fig:fig0}. Alternatively, if $\rho(k)$ is uniform across the entire history, then we recover the well-studied Elephant Random Walk~\cite{schutz2004elephants,coletti2017central,laulin2022introducing,gut2023elephant}. 

\begin{table}[b]
    \centering
    \includegraphics[width =\linewidth]{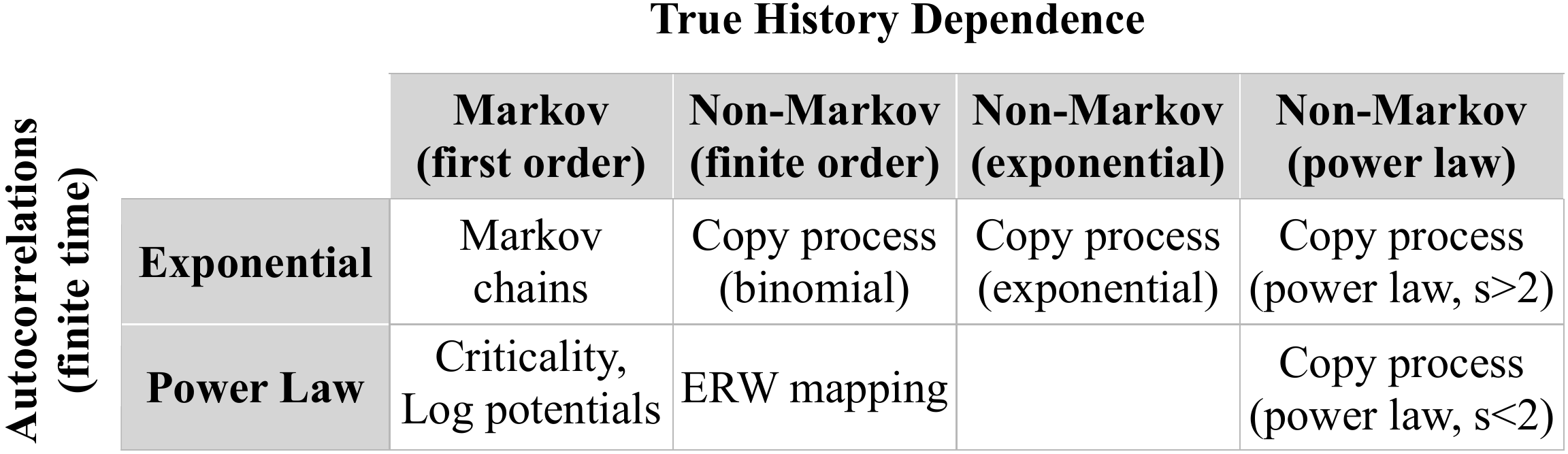}
    \caption{Examples of dynamical processes categorized by their history dependence (columns) and autocorrelation scaling (rows) that can be observed up to arbitrarily long times.}
    \label{tab:tab1}
\end{table}

For any finite-order history dependence [e.g., binomial $\rho(k)$ of order $N = 10$], the long-time autocorrelations decay exponentially, as expected [Fig.~\ref{fig:fig2}(c)]. Similarly, exponentially-decaying dependencies $\rho(k)$ produce exponentially-decaying autocorrelations for all $k$. For power-law history dependence $\rho(k) \propto k^{-s}$, the picture becomes more nuanced. For exponents $1 < s<2$, autocorrelations always follow a power law. However, for $s>2$, the autocorrelations display two qualitatively distinct regimes: exponential scaling up to a threshold $k^*$ and power-law scaling beyond this point~\cite{leighton2025companion}. Remarkably, this shows that even with power-law history dependence, autocorrelations can still decay exponentially over a range of times that can be made arbitrarily large by the appropriate tuning of $s$ and $\alpha$. In practice, this means that observing exponentially-decaying autocorrelations in finite data is insufficient to distinguish between finite-order, exponential, and power-law history dependence [Fig.~\ref{fig:fig2}(c)]. Table~\ref{tab:tab1} summarizes these results.

In stark contrast, the dynamical information correctly captures the true history dependence for all choices of $\rho(k)$ discussed above [Fig.~\ref{fig:fig2}(d)]. Under mild conditions on $\rho$, the dynamical information scales asymptotically as
\begin{equation}
I_k \sim \frac{1}{2}(2\alpha-1)^2\rho(k)^2,
\end{equation}
thus quantitatively matching the scaling of dependencies~\cite{leighton2025companion}. This allows us to reliably distinguish between Markovian, finite order, and infinite-order dynamics with exponential and power-law history dependence.

\begin{figure*}[t]
    \centering
    \includegraphics[width =\linewidth]{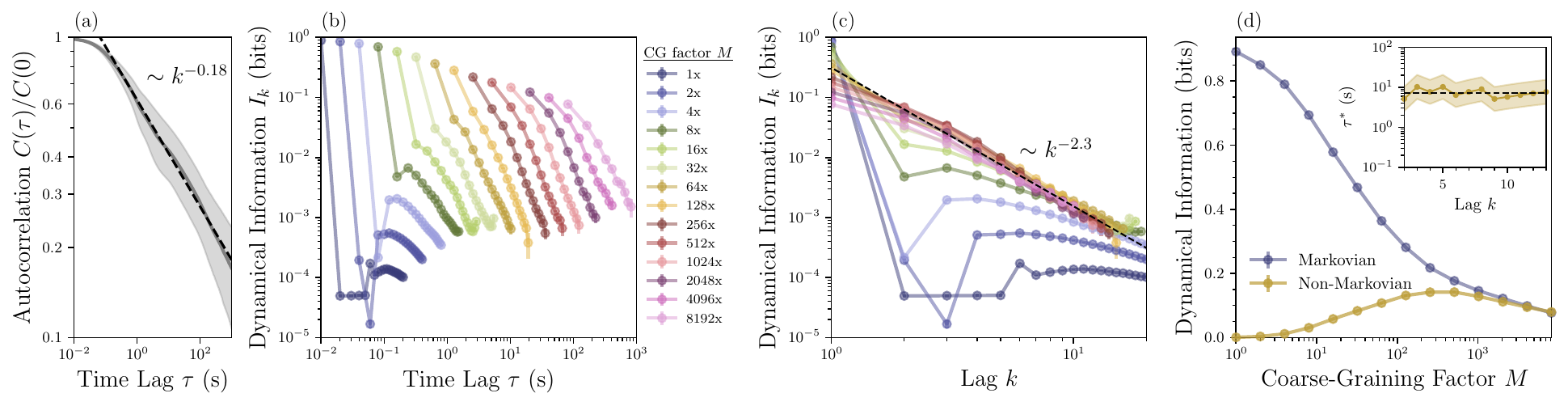}
    \caption{Non-Markovian dependencies in fly behavior. (a) Autocorrelations $C(\tau)$ as a function of time lag $\tau = k\Delta t$. Solid line: mean across all flies. Shaded region: one standard deviation. Dashed line: power-law fit. (b) Dynamical information $I_k$ versus time lag $\tau$ for different levels of temporal coarse-graining $M$. (c) As a function of the dimensionless lag $k$, dynamical information collapses onto a single curve. Dashed line: power-law fit. (d) Markovian ($I_1$) and non-Markovian ($I_{>1} = \sum_{k>1}I_k$, here summed to $k=10$) information versus coarse-graining factor $M$. Inset: For each order $k$, we compute the timescale $\tau^*$ at which $I_k$ is maximized (Fig.~\ref{fig:SI_Fig_OrdervsCG}). Gold shaded region: maximum and minimum possible $\tau^*$ values. Horizontal dashed line: mean value $\tau^* = 7.4$s.}
    \label{fig:fig3}
\end{figure*}

\emph{Inferring historical dependencies in fly behavior}.---As animals move and interact with their environment, they exhibit patterns of behavior across a wide range of time scales~\cite{alba2020exploring, bialek2024long, barabasi2005origin, wieczynski2025long}. Behavior, and that of the fly in particular, is characterized by long-range correlations~\cite{berman2016predictability, alba2020exploring, bialek2024long, stephens2008dimensionality, wiltschko2015mapping, barabasi2005origin}. In one view, these correlations emerge from short-range interactions among the many underlying degrees of freedom, perhaps near criticality~\cite{bak2013nature}. In another, they reflect truly long-range non-Markovian dependencies between behavioral states~\cite{alba2020exploring,berman2016predictability}. To distinguish between these pictures, we must quantify dependencies directly.

Experimental advances enable prolonged recordings of animal behavior with fine temporal resolution. We consider the behavior of 45 fruit flies, each recorded continuously for several days with resolution $\Delta t = 0.01$s in prior experiments~\cite{mckenzie2025capturing}. The behavior is coarse-grained into meaningful discrete states~\cite{mckenzie2025capturing, berman2016predictability, alba2020exploring}, which we further coarse-grain to two states that reflect activity ($x_t = 1$) or inactivity ($x_t = 0$; see Appendix). The autocorrelations decay as a power law over five orders of magnitude in time [Fig.~\ref{fig:fig3}(a)], with exponent $\sim0.18$ consistent with prior work~\cite{bialek2024long}. Yet, as discussed above, this is insufficient to draw conclusions about the underlying dependencies; for this, we turn to the dynamical information.

Due to the exponential explosion of possible trajectories with $k$, estimating information-theoretic quantities like $I_k$ from finite data is notoriously difficult~\cite{strong1998entropy}. However, given the significant length of the experiments, we are able to reliably estimate the dynamical information $I_k$ up to $k\approx 20$ using finite-data corrections (see Appendix). To explore even longer-range dependencies, we coarse-grain in time. Specifically, we partition each recording into bins of $M$ consecutive timepoints and randomly select one state within each bin, resulting in coarse-grained dynamics with steps of length $M\Delta t$ (see Appendix). The coarse-graining factor $M$ defines the timescale of the dynamics. For the largest factor $M = 8192$, we can reliably estimate $I_k$ out to $k = 12$, ultimately quantifying non-Markovian dependencies up to 16 minutes into the past.

As a function of time $\tau = kM\Delta t$, the autocorrelations are invariant under this coarse-graining by definition, thus collapsing to one curve [Fig.~\ref{fig:fig3}(a)]. The dynamical information, by contrast, varies significantly with the level of temporal coarse-graining [Fig.~\ref{fig:fig3}(b)]. For fine-grained dynamics (small $M$), the information $I_k$ is non-monotonic, first decreasing dramatically before reaching a local maximum at intermediate order $k$. This steep drop in $I_k$ arises because state transitions are extremely rare on short timescales $\tau$, yielding fine-grained dynamics that appear nearly Markovian. Coarse-graining the dynamics (increasing $M$) decreases the Markovian information $I_1$ while increasing the non-Markovian contributions $I_{k>1}$, leading to stronger historical dependencies. Moreover, the dynamical information approaches a consistent monotonic scaling [Fig.~\ref{fig:fig3}(b)], with dependencies of higher order $k$ providing less information.

To better understand this scaling, we plot the dynamical information as a function of the dimensionless lag $k$. Surprisingly, after only a few iterations of coarse-graining, the information $I_k$ collapses onto a single curve [Fig.~\ref{fig:fig3}(c)]. For over two orders of magnitude in scale $M$, the dynamical information appears to follow a power law $I_k \sim k^{-\gamma}$ (with $\gamma \approx 2.3$). Unlike the autocorrelations, this collapse does not follow necessarily from the definition of $I_k$. Instead, it suggests that non-Markovian dependencies in fly behavior are self-similar, maintaining the same functional form across timescales from fractions of a second to minutes.

Despite this timescale invariance, one might expect the overall strength of non-Markovian dependencies to depend on the scale $M$. For small $M$ the sparsity of state transitions yields dynamics that appear nearly Markovian, while in the limit of large $M$, coarse-grained dynamics must eventually lose all history dependence. This suggests the existence of a special timescale on which non-Markovian dependencies are strongest. To test this hypothesis, we compare the Markovian information ($I_1$) and total non-Markovian information ($I_{>1}=\sum_{k>1}I_k$) over a range of coarse-graining factors $M$ [Fig.~\ref{fig:fig3}(d)]. As expected, in the fine-grained limit higher-order dependencies vanish ($I_{>1}\approx0$), making the dynamics nearly Markovian ($I_\text{tot} \approx I_1$). As the timescale increases, the Markovian information decreases while the non-Markovian information increases, reaching a peak at intermediate $M$. Thus, temporal coarse-graining induces non-Markovianity. After further coarse-graining, the total dynamical information $I_\text{tot}$ decays to zero as the future becomes decoupled from the past.

The peak in the total non-Markovian information $I_{>1}$ also arises at each individual order $I_k$ [Fig.~\ref{fig:SI_Fig_OrdervsCG}(a)]. Thus, for each order $k$, the dynamics define a timescale $\tau^*$ on which the $k^\text{th}$-order dependence is strongest [Fig.~\ref{fig:SI_Fig_OrdervsCG}(b)]. Remarkably, we find that this timescale is nearly constant across all orders $k$ [Fig.~\ref{fig:fig3}(d), \textit{inset}]. This implies that there is a unique time lag at which all non-Markovian dependencies are maximized. Notably, the identified timescale (around $7$ seconds) is similar to other key timescales in fly behavior; namely, those of working memory and burstiness~\cite{neuser2008analysis, kuntz2017visual, sorribes2011origin}. This suggests that the strongest historical dependencies in behavior may be driven by memory, which is an inherently non-Markovian phenomenon. Together, these results indicate that long timescales in fly behavior emerge from underlying non-Markovian dependencies that are (i) invariant in form, yet (ii) strongest on the scale of seconds.

\emph{Discussion}.---Biology features a plethora of stochastic processes with complex historical dependencies~\cite{vroylandt2022likelihood, d2014mechanisms, zeraati2023intrinsic, cavanagh2020diversity, elgamel2023multigenerational, alba2020exploring, bialek2024long, barabasi2005origin, shannon1948mathematical, wieczynski2025long}. In this Letter, we show that strength of non-Markovian dependencies can be quantified and decomposed using tools from information theory. The past reduces one's uncertainty about the future by an amount $I_\text{tot}$, which we refer to as the dynamical information. Each $k^\text{th}$-order dependence makes a non-negative contribution $I_k$ to this information, leading to a full decomposition $I_\text{tot} = I_1 + I_2 + \cdots$. In minimal models of non-Markovian dynamics the dynamical information captures the underlying nature of dependencies, even when autocorrelations categorically do not (Figs.~\ref{fig:fig0} and \ref{fig:fig2}). Applying this framework to recordings of fly behavior, we discover that the scaling of non-Markovian dependencies is invariant across two orders of magnitude in time [Fig.~\ref{fig:fig3}(c)]. The overall strength of dependencies, however, is maximized on the scale of seconds, thus identifying a unique timescale on which fly behavior is least Markovian [Fig.~\ref{fig:fig3}(d)].

Moving forward, an important practical challenge is quantifying long-range dependencies in biological data. Calculating the dynamical information $I_k$ is data intensive, requiring exponentially more data with larger $k$. As the length and size of biological experiments continues to grow, we anticipate that longer-range dependencies will become accessible. Future theoretical work could also seek to develop scalable techniques for bounding or approximating the dynamical information $I_k$, thereby providing a window deeper into the past.

We began this Letter with the observation that non-Markovian dynamics fundamentally arise through coarse-graining. In the context of fly behavior, we showed explicitly how coarse-graining in time can increase the strength of historical dependencies. Our framework is general---it does not rely on model assumptions nor specific biological details---and can thus be applied directly to other living systems across scales. This opens the door for future investigations into how non-Markovian dynamics emerge through coarse-graining in time and space.

\emph{Acknowledgements}---We thank Kevin Chen (Yale QBio) for feedback on the manuscript. This work was supported in part by Mossman and NSERC Postdoctoral Fellowships (M.P.L.), and by support from the Department of Physics, Quantitative Biology Institute, and Wu Tsai Institute at Yale University (C.W.L).
\bibliography{main}

\section*{End Matter}
\section*{Appendix A: Dynamical Information is a Total Correlation Rate}

Here we show that the dynamical information can be written as the long-time growth rate of the total correlation. Consider a discrete-state discrete-time stochastic process that starts at time $t=1$ and converges towards a steady state as $t\to\infty$. The dynamical information is defined as
\begin{equation}
I_\mathrm{tot} = h_0 - h_\infty,
\end{equation}
for marginal entropy $h_0 = H[x_t]$ and entropy rate $h_\infty$, which can be written as~\cite{Cover2006_Elements}
\begin{equation}\label{eq:entropyrate}
h_\infty = \lim_{t\to \infty} \frac{1}{t}H[x_t,x_{t-1},...,x_1].
\end{equation}
Note that since, at steady state, the marginal entropy must converge to a time-independent value, it can also be written as a rate in the long-time limit as
\begin{equation}\label{eq:marginalentropy}
h_0 = \lim_{t\to\infty}\frac{1}{t}\sum_{k=0}^{t-1} H[x_{t-k}].
\end{equation}
Now consider the total correlation~\cite{crooks2017measures} between $x_t$ and the entire history of the process,
\begin{equation}
\mathrm{TC}[x_t,x_{t-1},...,x_1] \equiv \sum_{k=0}^{t-1} H[x_{t-k}] - H[x_t,x_{t-1},...,x_1].
\end{equation}
Dividing by $t$ and taking the limit $t\to\infty$, we see that the dynamical information is the long-time rate of increase of the total correlation:
\begin{equation}
I_\mathrm{tot} = \lim_{t\to\infty} \frac{1}{t}\left( \sum_{k=0}^{t-1} H[x_{t-k}] - H[x_{t},x_{t-1},...,x_1]\right).
\end{equation}

\section*{Appendix B: Fly Behavior Analysis}

\subsection{Dataset}\label{app:dataset}

We explore history dependence in fly behavior using a publicly available dataset from Ref.~\cite{mckenzie2025capturing}, which captured time-courses of behavior for 47 flies over four to eight days with a resolution of $\Delta t = 0.01$s. Nine distinct behavioral states are resolved. We omit flies 7 and 45, for which data was not available, leaving a total of 45 flies for our analysis.

\subsection{Coarse-Graining Procedures}\label{app:cgprocedures}

To facilitate the estimation of dynamical information from finite data, we coarse-grain the original nine behavioral states into two states. Following Ref.~\cite{alba2020exploring}, we map each behavioral state to either `$x_t = 0$' (idle or unresolved) or `$x_t = 1$' (all other behaviors). 

\begin{figure}[h]
    \centering
    \includegraphics[width =0.8\linewidth]{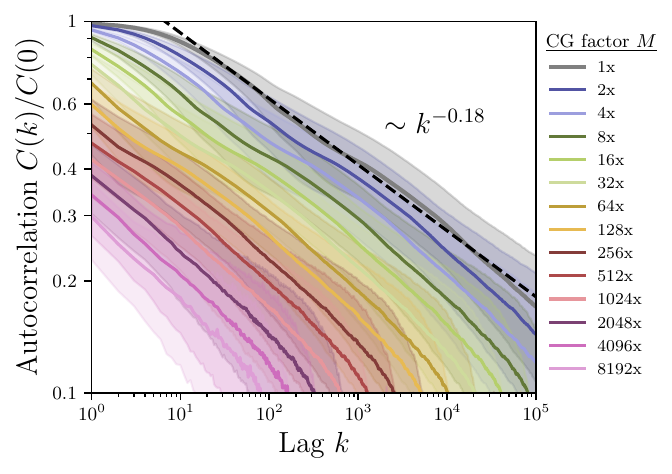}
    \caption{Autocorrelation $C(k)$ versus lag $k$ for different levels of temporal coarse-graining. Dashed line: power-law fit for the fine-grained autocorrelations ($M = 1$).}
    \label{fig:SI_Fig_autocorrelations}
\end{figure}

To investigate dependencies across different timescales, we also coarse-grain the dynamics in time. For a given coarse-graining factor $M$, we bin together $M$ consecutive timepoints and randomly select one state from each bin. This coarse-graining procedure ensures that the autocorrelations are invariant as a function of time $\tau = kM\Delta t$ [Fig.~\ref{fig:fig3}(a)], thus preserving the statistical relationships between past and present. When plotted versus the dimensionless lag $k$, the autocorrelations separate for different coarse-graining factors $M$ while maintaining the same power-law scaling (Fig.~\ref{fig:SI_Fig_autocorrelations}).

\subsection{Finite-Data Corrections}

When estimating information-theoretic quantities (and mutual information in particular), finite-data effects must be taken into account. When computing the dynamical information, we correct for finite-data effects using the methodology in Ref.~\cite{strong1998entropy}. For each lag $k$ and coarse-graining factor $M$, we compute the dynamical information $I_k$ using different fractions $f$ of the flies in the dataset (sampling many subsets for each $f$ to obtain estimates with uncertainties). We then fit a quadratic polynomial to the estimates of $I_k$ versus the inverse data fraction $1/f$. The intercept of the resulting fit at $1/f = 0$ is our infinite-data estimate for the dynamical information.

We also perform the same analysis after shuffling the dynamics for each fly in time. This destroys all temporal dependencies and therefore should yield an estimate of $I_k=0$ for all $k$ and coarse-graining factors $M$. For any $k$ and $M$ where the shuffled $I_k$ is not within uncertainty of zero, we consider the finite-data estimates unreliable. For each $M$, this sets the upper bound on $k$ for which we can estimate $I_k$.

\subsection{Power-Law Fits}\label{app:powerlaws}

For the autocorrelations in Fig.~\ref{fig:fig3}(a), we fit a power law to the data for $0.1\mathrm{s}\leq \tau\leq 1000$s. We obtain an exponent of $0.178$ with a standard error of $0.001$ ($R^2 = 1.00$). We also fit a power law to the dynamical information $I_k$ as a function of dimensionless lag $k$ [Fig.~\ref{fig:fig3}(c)] by combining all datapoints with $k\geq 3$ and coarse-graining factor $M \geq 32$. This yields an exponent of $\gamma = 2.31\pm 0.04$ (standard error)
with $R^2=0.97$.

\begin{figure}[h]
    \centering
    \includegraphics[width =0.8\linewidth]{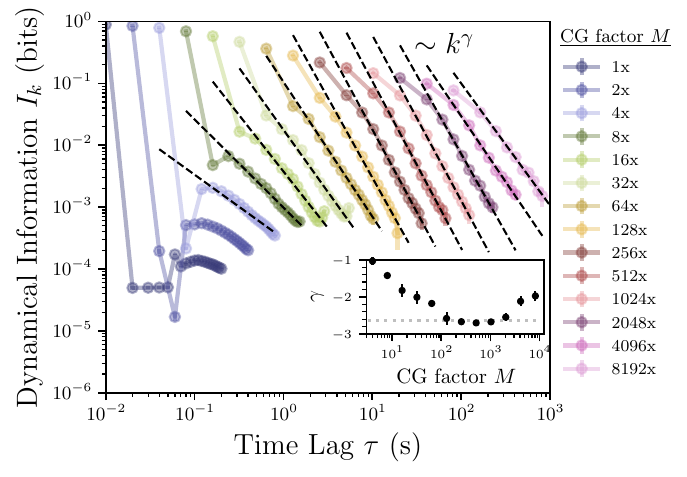}
    \caption{Dynamical information $I_k$ versus time lag $\tau$ for different levels of temporal coarse-graining $M$. Dashed lines: power-law fits for $k\geq3$. Inset: power-law exponents $\gamma$ versus coarse-graining factor $M$ with error bars reflecting standard errors and dashed line indicating the mean of the five largest exponents.}
    \label{fig:SI_Fig_PowerLaws}
\end{figure}

Finally, we fit power laws to the dynamical information as a function of dimensional time $\tau = kM\Delta t$ for each coarse-graining factor $M \ge 4$ and lag $k\geq 3$. Figure~\ref{fig:SI_Fig_PowerLaws} shows the resulting power-law fits, with the exponents shown in the inset. For coarse-graining factors between $128$ and $2048$ (inclusive), we obtain an approximately constant scaling exponent of $\gamma \approx 2.6$. 

\begin{figure}[h]
    \centering
    \includegraphics[width =\linewidth]{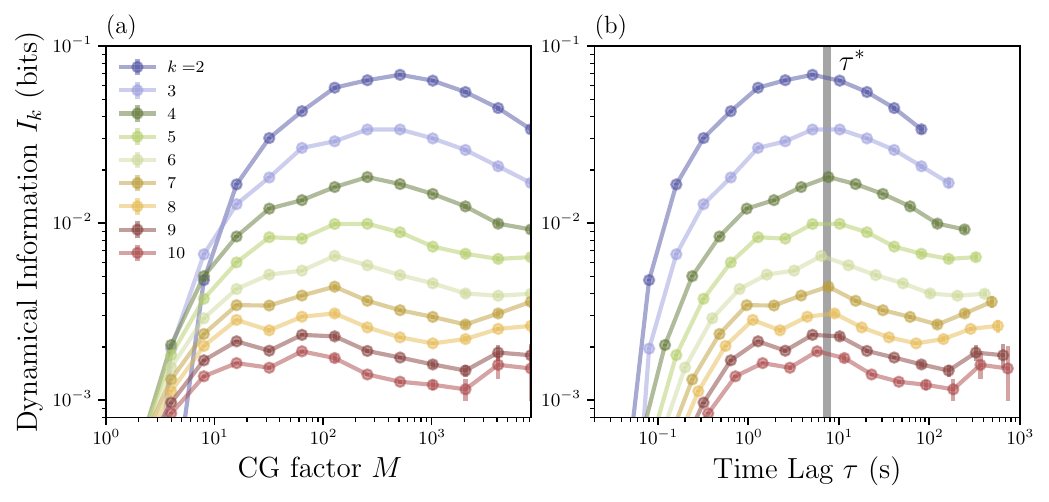}
    \caption{Dynamical information $I_k$ of different orders $k$ plotted versus (a) the coarse-graining factor $M$, and (b) the corresponding timescale $\tau = kM\Delta t$. Vertical line indicates the timescale $\tau^* \approx 7.4$s on which dynamical information is maximized for all orders $k$.}
    \label{fig:SI_Fig_OrdervsCG}
\end{figure}

\subsection{Timescale Calculations}\label{app:timescales}

Figure~\ref{fig:SI_Fig_OrdervsCG} shows the $k^\mathrm{th}$-order dynamical information $I_k$ for orders $2\leq k\leq10$ as functions of the coarse-graining factor $M$. Notably, each curve exhibits a maximum at an intermediate level of coarse-graining [Fig.~\ref{fig:SI_Fig_OrdervsCG}(a)]. Plotting $I_k$ as a function of time lag $\tau = kM\Delta t$ for different $M$, we find that the maxima align at a specific timescale $\tau^*$ [Fig.~\ref{fig:SI_Fig_OrdervsCG}(b)]. For each order $k$, we plot the timescale $\tau^*$ at which $I_k$ is maximized in the inset of Fig.~\ref{fig:fig3}(d), with uncertainties arising from the finite resolution of coarse-graining factors. Within uncertainty, these timescales all appear to be identical, with a mean value across all $k$ of $\tau^*\approx 7.4$s.

\subsection{Data and Code Availability}\label{app:dataandcode}
All experimental data used in this Letter was obtained in Ref.~\cite{mckenzie2025capturing}, and is publicly available at Ref.~\cite{mckenzesmith_grace_c_2023}. Our code for analyzing this data and producing Figs.~\ref{fig:fig3}-\ref{fig:SI_Fig_OrdervsCG} is publicly available at Ref.~\cite{github}.

\end{document}